\documentclass[12pt,letter]{article}

\usepackage[top=20truemm,bottom=20truemm,left=20truemm,right=20truemm]{geometry}

\usepackage[dvips]{graphicx,graphics}
\usepackage[cmex10]{amsmath}
\usepackage{amssymb} 
\usepackage{amsthm}
\newtheorem{definition}{Definition}
\newtheorem{theorem}{Theorem}
\newtheorem*{remark}{Remark}

\title{\LARGE \bf
	Simple Design on Nanoscale Receivers Using CNT Cantilevers
}

\author{Yuji Ito%
	\thanks{Yuji Ito and Yukihiro Tadokoro are with TOYOTA CENTRAL R\&D LABS., INC., 41-1 Yokomichi, Nagakute-shi, Aichi 480-1192, Japan}
	\thanks{Corresponding Author: {\tt\small ito-yuji@mosk.tytlabs.co.jp} } 	
	\and
	Yukihiro Tadokoro$^{\ast}$
}

\date{\today}

\begin{document}

\maketitle

%%%%%%%%%%%%%%%%%%%%%%%%%%%%%%%%%%%%%%%%%%%%%%%%%%%%%%%%%%%%%%%%%%%%%%%%%%%%%%%%%%%%%%%%%%%%%%%%%%%%%%%%%%%%%%%%%%%%%%%%%%%%%%%%%%%%%%%
%%%%%%%%%%%%%%%%%%%%%%%%%%%%%%%%%%%%%%%%%%%%%%%%%%%%%%%%%%%%%%%%%%%%%%%%%%%%%%%%%%%%%%%%%%%%%%%%%%%%%%%%%%%%%%%%%%%%%%%%%%%%%%%%%%%%%%%
%%%%%%%%%%%%%%%%%%%%%%%%%%%%%%%%%%%%%%%%%%%%%%%%%%%%%%%%%%%%%%%%%%%%%%%%%%%%%%%%%%%%%%%%%%%%%%%%%%%%%%%%%%%%%%%%%%%%%%%%%%%%%%%%%%%%%%%
\iffalse\section{symbols}\fi

%%%%%%%%%%%%%%%%%%%%%%%%%%%%%%%%%%%%%%
\newcommand{\Wave}{E}
\newcommand{\Mag}{A}%\tilde{E}}
\newcommand{\Freq}{\omega}
\newcommand{\Bias}{B}%\overline{E}}
\newcommand{\Phase}{\theta}
%%%%%%%%%%%%%%%%%%%%%%%%%%%%%%%%%%%%%%

\newcommand{\CNTmass}{{m}}

\newcommand{\charge}{{q}}
\newcommand{\elasticity}{{k}}
\newcommand{\viscosity}{{\gamma}}

\newcommand{\Voltage}{V}

\newcommand{\Disp}{x}

\newcommand{\SSDisp}{x_{\mathrm{ss}}}
\newcommand{\SSDispMag}{\tilde{x}_{\mathrm{ss}}}
\newcommand{\SSDispPhase}{\Phase_{\mathrm{ss}}}

\newcommand{\SSestDispPhase}{\hat{\Phase}_{\mathrm{ss}}}

\newcommand{\InWave}{\Wave_{\mathrm{in}}}
\newcommand{\InMag}{\Mag_{\mathrm{in}}}
\newcommand{\InFreq}{\Freq_{\mathrm{in}}}
\newcommand{\InPhase}{\Phase_{\mathrm{in}}}

\newcommand{\RefWave}{\Wave_{\mathrm{r}}}
\newcommand{\RefMag}{\Mag_{\mathrm{r}}}
\newcommand{\RefFreq}{\Freq_{\mathrm{r}}}

\newcommand{\CarrierSig}{f_{\mathrm{c}}}
\newcommand{\SymbolTime}[1]{T_{#1}}

\newcommand{\SymbA}{+}%\mathrm{a}}
\newcommand{\SymbB}{-}%\mathrm{b}}

\newcommand{\Current}{I}
\newcommand{\ConstCurrent}{I_{0}}
\newcommand{\LinearCurrent}{I_{1}}

\newcommand{\CurNoise}{e}
\newcommand{\CurIntNoise}{n_{e}}

\newcommand{\DemoduSig}{D}

\newcommand{\ObjectiveFunc}{J}

%%%%%%%%%%%%%%%%%%%%%%%%%%%%%%%%%%%%%%%%%%%%%%%%%%%%%%%%%%%%%%%%%%%%%%%%%%%%%%%%%%%%%%%%%%%%%%%%%%%%%%%%%%%%%%%%%%%%%%%%%%%%%%%%%%%%%%%
%%%%%%%%%%%%%%%%%%%%%%%%%%%%%%%%%%%%%%%%%%%%%%%%%%%%%%%%%%%%%%%%%%%%%%%%%%%%%%%%%%%%%%%%%%%%%%%%%%%%%%%%%%%%%%%%%%%%%%%%%%%%%%%%%%%%%%%
%%%%%%%%%%%%%%%%%%%%%%%%%%%%%%%%%%%%%%%%%%%%%%%%%%%%%%%%%%%%%%%%%%%%%%%%%%%%%%%%%%%%%%%%%%%%%%%%%%%%%%%%%%%%%%%%%%%%%%%%%%%%%%%%%%%%%%%
\begin{abstract}
A nanoscale receiver utilizing the cantilever of a carbon nanotube has been developed to detect phase information included in transmitted signals.
The existing receiver consists of a phase detector and demodulator which employ a reference wave and carrier signal, respectively.
This paper presents a design method to simplify the receiver in structure with enhancing the performance for the phase detection. 
The reference wave or carrier signal is not needed in the receiver via the proposed design method.
\end{abstract}

%%%%%%%%%%%%%%%%%%%%%%%%%%%%%%%%%%%%%%%%%%%%%%%%%%%%%%%%%%%%%%%%%%%%%%%%%%%%%%%%%%%%%%%%%%%%%%%%%%%%%%%%%%%%%%%%%%%%%%%%%%%%%%%%%%%%%%%
%%%%%%%%%%%%%%%%%%%%%%%%%%%%%%%%%%%%%%%%%%%%%%%%%%%%%%%%%%%%%%%%%%%%%%%%%%%%%%%%%%%%%%%%%%%%%%%%%%%%%%%%%%%%%%%%%%%%%%%%%%%%%%%%%%%%%%%
%%%%%%%%%%%%%%%%%%%%%%%%%%%%%%%%%%%%%%%%%%%%%%%%%%%%%%%%%%%%%%%%%%%%%%%%%%%%%%%%%%%%%%%%%%%%%%%%%%%%%%%%%%%%%%%%%%%%%%%%%%%%%%%%%%%%%%%
\section{Introduction}

Nanoscale sensors have the potential to realize next-generation sensing systems.
An interesting example is smartdust to measure physical quantities such as temperature and light via tiny sensors \cite{Kahn99}.
Communication between nanoscale sensors is essential to transmit and aggregate the measured quantities.
Unfortunately, traditional electromagnetic-based antennas cannot be applied to nanoscale networks because the antenna size is on the order of the wavelengths of transmitted signals \cite{Stutzman81}, e.g., the size of an antenna for the megahertz band is several centimeters.
Although using signals in the terahertz band may downsize antennas \cite{Jornet12a,Jornet12b,Sugiura14}, it requires an additional cost to develop circuits and systems.

Nanoscale receivers have been developed with the antennas using the cantilevers of carbon nanotubes (CNTs) \cite{Jensen07,Vincent11, Tadokoro2018}.
Underlying technology is nanomechanical resonator, where physical quantities can be measured through observing mechanical vibration \cite{Ayari2007,Moser2014}.
The nanoscale receiver presented in \cite{Tadokoro2018} can detect the phase information of an incoming electromagnetic wave.
Unfortunately, this receiver requires an additional reference electromagnetic wave and a carrier signal for the detection.
The structure of the receiver is complicated by implementing devices to generate such a wave and signal.
The advantage of the receiver, i.e., being infinitesimal, may be degraded.

To overcome this problem, this paper proposes a design method which simplifies the structure of the nanoscale receiver in \cite{Tadokoro2018}. 
A performance measure for the phase detection is formulated as a function of the motion of the CNT.
A decomposition method factorizes the performance measure function.
Analyzing the factorized function designs parameters in the receiver such that either the reference wave or the carrier signal is excluded from the receiver. 
The receiver without the reference wave or the carrier signal is simplified in structure.
The effectiveness of the proposed design method is demonstrated through a numerical example.

%%%%%%%%%%%%%%%%%%%%%%%%%%%%%%%%%%%%%%%%%%%%%%%%%%%%%%%%%%%%%%%%%%%%%%%%%%%%%%%%%%%%%%%%%%%%%%%%%%%%%%%%%%%%%%%%%%%%%%%%%%%%%%%%%%%%%%%
%%%%%%%%%%%%%%%%%%%%%%%%%%%%%%%%%%%%%%%%%%%%%%%%%%%%%%%%%%%%%%%%%%%%%%%%%%%%%%%%%%%%%%%%%%%%%%%%%%%%%%%%%%%%%%%%%%%%%%%%%%%%%%%%%%%%%%%
%%%%%%%%%%%%%%%%%%%%%%%%%%%%%%%%%%%%%%%%%%%%%%%%%%%%%%%%%%%%%%%%%%%%%%%%%%%%%%%%%%%%%%%%%%%%%%%%%%%%%%%%%%%%%%%%%%%%%%%%%%%%%%%%%%%%%%%
\section{System configuration \label{sec_config}}

\begin{figure}[t]
	\centering
	\includegraphics[width=6.0in]{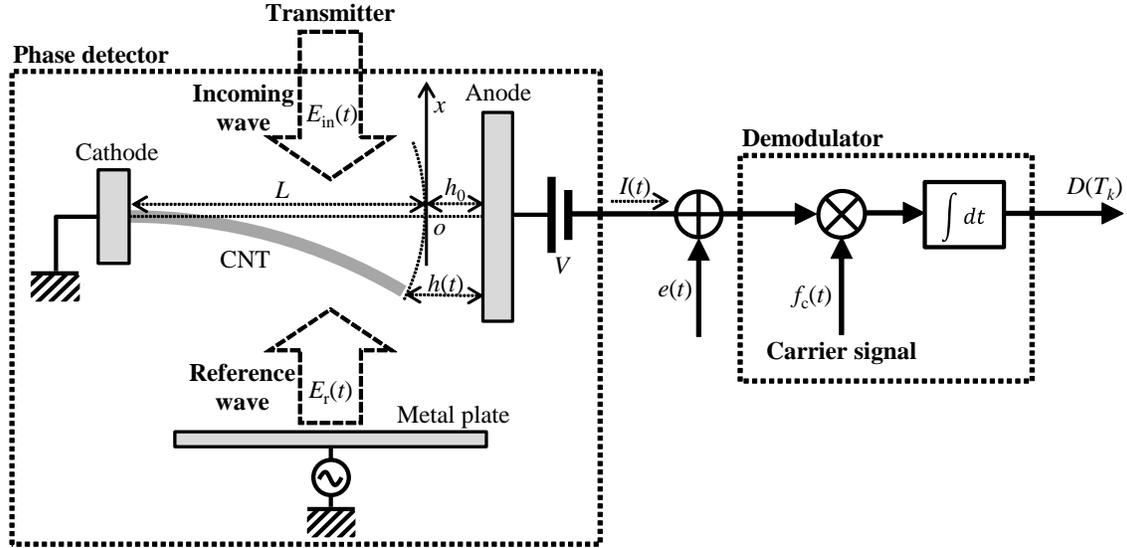}
	\caption{Configuration of the Nanoscale receiver. 
	In the proposed design method using Theorems \ref{thm:wt_Corr} and \ref{thm:wt_RefWave}, either the reference wave or the carrier signal can be omitted.}\label{fig:sys_config}
\end{figure}

This section reviews the configuration of the nanoscale receiver proposed in \cite{Tadokoro2018}.
The receiver obtains the information of the phase $\InPhase$ of the incoming wave $\InWave$ which is sent from the transmitter.
The receiver consists of a phase detector and demodulator as shown in Fig. \ref{fig:sys_config}.
A CNT is arrayed on the cathode connected to the ground in the phase detector. 
Applying the voltage $\Voltage$ to the anode excites a charge around the tip of the CNT.
The tip of the CNT is subject to an electric force which is generated by the charge, the incoming wave $\InWave$, and the reference wave $\RefWave$ according to the Coulomb's law. 
The electric force is sinusoidal because the incoming wave $\InWave$ is a cosine wave with the offset $\InPhase$.
Such a sinusoidal force vibrates the tip of the CNT, depending on the phase $\InPhase$.
Meanwhile, a field emission current flows from the tip of the CNT to the anode which is generated by the voltage $\Voltage$.
The time-series of the current depends on the vibration of the tip, i.e., the phase $\InPhase$.
The demodulator extracts the information of the phase $\InPhase$ from the current under an appropriate setting of the carrier signal.
Mathematical models used in the nanoscale receiver are introduced in Sections \ref{sec_incoming_wave}--\ref{sec_demodulator}.

%%%%%%%%%%%%%%%%%%%%%%%%%%%%%%%%%%%%%%%%%%%%%%%%%%%%%%%%%%%%%%%%%%%%%%%%%%%%%%%%%%%%%%%%%%%%%%%%%%%%%%%%%%%%%%%%%%%%%%%%%%%%%%%%%%%%%%%
%%%%%%%%%%%%%%%%%%%%%%%%%%%%%%%%%%%%%%%%%%%%%%%%%%%%%%%%%%%%%%%%%%%%%%%%%%%%%%%%%%%%%%%%%%%%%%%%%%%%%%%%%%%%%%%%%%%%%%%%%%%%%%%%%%%%%%%
%%%%%%%%%%%%%%%%%%%%%%%%%%%%%%%%%%%%%%%%%%%%%%%%%%%%%%%%%%%%%%%%%%%%%%%%%%%%%%%%%%%%%%%%%%%%%%%%%%%%%%%%%%%%%%%%%%%%%%%%%%%%%%%%%%%%%%%
\subsection{Incoming wave \label{sec_incoming_wave}}

This subsection reviews the incoming wave \cite{Tadokoro2018} which is sent from the transmitter.
The incoming wave $\InWave(t)$ at the time $t$ is defined as
\begin{equation} 
	\begin{aligned}
		& 
		\InWave(t) := \InMag \cos ( \InFreq t + \InPhase  ) 
		,
	\end{aligned}
\end{equation}
where $\InPhase \in \{ \InPhase^{\SymbA}, \InPhase^{\SymbB} \}$.
The symbols $\InPhase^{\SymbA}$ and $\InPhase^{\SymbB}$ are the phases corresponding to one-bit signals, respectively, which are arbitrary designed.
The receiver attempts to distinguish between the two types of the phases. 
In the following, the superscripts $(\cdot)^{\SymbA}$ and $(\cdot)^{\SymbB}$ denote variables/functions corresponding to $\InPhase^{\SymbA}$ and $\InPhase^{\SymbB}$.

%%%%%%%%%%%%%%%%%%%%%%%%%%%%%%%%%%%%%%%%%%%%%%%%%%%%%%%%%%%%%%%%%%%%%%%%%%%%%%%%%%%%%%%%%%%%%%%%%%%%%%%%%%%%%%%%%%%%%%%%%%%%%%%%%%%%%%%
%%%%%%%%%%%%%%%%%%%%%%%%%%%%%%%%%%%%%%%%%%%%%%%%%%%%%%%%%%%%%%%%%%%%%%%%%%%%%%%%%%%%%%%%%%%%%%%%%%%%%%%%%%%%%%%%%%%%%%%%%%%%%%%%%%%%%%%
%%%%%%%%%%%%%%%%%%%%%%%%%%%%%%%%%%%%%%%%%%%%%%%%%%%%%%%%%%%%%%%%%%%%%%%%%%%%%%%%%%%%%%%%%%%%%%%%%%%%%%%%%%%%%%%%%%%%%%%%%%%%%%%%%%%%%%%
\subsection{Phase detector \label{sec_phase_detector}}

This subsection reviews the phase detector \cite{Tadokoro2018} which converts the motion of CNT's tip into the field emission current.
The motion equation with respect to the displacement $\Disp(t)$ of the CNT's tip at the time $t$ is modeled as the liner equation: 
\begin{align}
	\CNTmass \frac{\mathrm{d}^{2}{\Disp(t)}}{\mathrm{d} t^{2}}  + \viscosity \frac{\mathrm{d}{\Disp(t)}}{\mathrm{d} t}  +  \elasticity \Disp(t) & =  \charge (\InWave(t) - \RefWave(t))
	,
	\label{eq:CNT_motion}
\end{align}
where $\charge$, $\CNTmass$, $\viscosity$, and $\elasticity$ are the amount of the charge around the tip, effective mass, viscosity, and elasticity, respectively.
Assuming that the displacement $\Disp(t)$ of the CNT's tip is sufficiently small, the field emission current $\Current(t)$ is approximately a function of $\Disp(t)$:
	\begin{equation} \label{eq:current}
		\begin{aligned}
			\Current(t)& 
			\approx \ConstCurrent +  \LinearCurrent \Disp(t)^{2} 
			,
		\end{aligned}
	\end{equation}
	where $\ConstCurrent$ and $\LinearCurrent$ are constants.

%%%%%%%%%%%%%%%%%%%%%%%%%%%%%%%%%%%%%%%%%%%%%%%%%%%%%%%%%%%%%%%%%%%%%%%%%%%%%%%%%%%%%%%%%%%%%%%%%%%%%%%%%%%%%%%%%%%%%%%%%%%%%%%%%%%%%%%
%%%%%%%%%%%%%%%%%%%%%%%%%%%%%%%%%%%%%%%%%%%%%%%%%%%%%%%%%%%%%%%%%%%%%%%%%%%%%%%%%%%%%%%%%%%%%%%%%%%%%%%%%%%%%%%%%%%%%%%%%%%%%%%%%%%%%%%
%%%%%%%%%%%%%%%%%%%%%%%%%%%%%%%%%%%%%%%%%%%%%%%%%%%%%%%%%%%%%%%%%%%%%%%%%%%%%%%%%%%%%%%%%%%%%%%%%%%%%%%%%%%%%%%%%%%%%%%%%%%%%%%%%%%%%%%
\subsection{Demodulator \label{sec_demodulator}}

This subsection reviews the demodulator \cite{Tadokoro2018} which extracts the phase information from the field emission current.
Let us define the symbol duration $\SymbolTime{s}$ over which the field emission current $\Current(t)$ is integrated:
\begin{equation} \label{eq:def_time_horizon}
\begin{aligned}
&
\SymbolTime{s} := \frac{2\pi}{\InFreq}  s
,
\end{aligned}
\end{equation}
where $s \in \mathbb{N}$ denotes the number of the periods.
The demodulator combines the field emission current $\Current(t)$ with the carrier signal $\CarrierSig(t)$ and integrates it subject to the noise $\CurNoise(t)$:
	\begin{equation} \label{eq:def_DemoduSig}
		\begin{aligned}
			&
			\DemoduSig(\SymbolTime{s})
			:=
			\frac{1}{\SymbolTime{s}}
			\int_{0}^{\SymbolTime{s}}
			(\Current(t) + \CurNoise(t) ) \CarrierSig(t)
			\mathrm{d} t
			=
			\DemoduSig_{0}(\SymbolTime{s}) + \CurIntNoise(\SymbolTime{s})
			,
		\end{aligned}
	\end{equation}
where $\DemoduSig_{0}(\SymbolTime{s}):=\DemoduSig(\SymbolTime{s})|_{\CurNoise(t)=0}$.
Using the signal $\DemoduSig(\SymbolTime{s})$ determines the estimation $\hat{\InPhase}$ of $\InPhase$ between $\InPhase^{\SymbA}$ and $\InPhase^{\SymbB}$ under the assumption that we know the fact whether $\DemoduSig_{0}^{\SymbA}(\SymbolTime{s}) < \DemoduSig_{0}^{\SymbB}(\SymbolTime{s})$ or $\DemoduSig_{0}^{\SymbA}(\SymbolTime{s}) > \DemoduSig_{0}^{\SymbB}(\SymbolTime{s})$.
	This assumption is reasonable if the incoming wave firstly transmits the phase $\InPhase^{\SymbA}$ for the initialization.
	The performance for the phase detection by the receiver depends on the distance between constellation points:
	\begin{equation} \label{eq:def_objective}
		\begin{aligned}
			\ObjectiveFunc(\SymbolTime{s})
			:=
			|\DemoduSig_{0}^{\SymbA}(\SymbolTime{s}) - \DemoduSig_{0}^{\SymbB}(\SymbolTime{s})|
			\;=
			\Big|
			\frac{\LinearCurrent}{\SymbolTime{s}}
			\int_{0}^{\SymbolTime{s}}
			( \Disp^{\SymbA}(t)^{2} - \Disp^{\SymbB}(t)^{2} ) \CarrierSig(t)
			\mathrm{d} t 
			\Big|
			,
		\end{aligned}
\end{equation}
where the above equality holds because of (\ref{eq:current}) and (\ref{eq:def_DemoduSig}).

%%%%%%%%%%%%%%%%%%%%%%%%%%%%%%%%%%%%%%%%%%%%%%%%%%%%%%%%%%%%%%%%%%%%%%%%%%%%%%%%%%%%%%%%%%%%%%%%%%%%%%%%%%%%%%%%%%%%%%%%%%%%%%%%%%%%%%%
%%%%%%%%%%%%%%%%%%%%%%%%%%%%%%%%%%%%%%%%%%%%%%%%%%%%%%%%%%%%%%%%%%%%%%%%%%%%%%%%%%%%%%%%%%%%%%%%%%%%%%%%%%%%%%%%%%%%%%%%%%%%%%%%%%%%%%%
%%%%%%%%%%%%%%%%%%%%%%%%%%%%%%%%%%%%%%%%%%%%%%%%%%%%%%%%%%%%%%%%%%%%%%%%%%%%%%%%%%%%%%%%%%%%%%%%%%%%%%%%%%%%%%%%%%%%%%%%%%%%%%%%%%%%%%%
\section{Problem setting \label{sec_problem}}

\newcommand{\SSObjectiveFunc}{J_{\mathrm{ss}}}

This paper attempts to simplify the structure of the receiver and to improve the performance for the phase detection simultaneously.
A reasonable strategy to enhance the performance is maximizing  $\ObjectiveFunc(\SymbolTime{s})$, which is termed the \textit{performance index} in the following.
Additionally, a constraint is introduced to keep the variance of the noise $\CurIntNoise(\SymbolTime{s})$ in (\ref{eq:def_DemoduSig}) constant for any $\CarrierSig(t)$:
\begin{equation} \label{eq:Corr_constraint}
	\begin{aligned}
		&
		(1/\SymbolTime{s}) \int_{0}^{\SymbolTime{s}}   \CarrierSig(t)^{2}  \mathrm{d} t
		=
		1
		.
	\end{aligned}
\end{equation}
We design $\RefWave(t)$, $\CarrierSig(t)$, $\InPhase^{\SymbA}$, and $\InPhase^{\SymbB}$ under the constraint (\ref{eq:Corr_constraint}) as follows.

\textbf{Main problem:} Design $\RefWave(t)$, $\CarrierSig(t)$, $\InPhase^{\SymbA}$, and $\InPhase^{\SymbB}$ such that the receiver is simplified in structure with maximizing the performance index, i.e.,
\begin{equation} \label{eq:main_problem}
	\begin{aligned}	
		\max_{\RefWave(t), \CarrierSig(t), \InPhase^{\SymbA},  \InPhase^{\SymbB}  }
		\ObjectiveFunc(\RefWave, \CarrierSig, \InPhase^{\SymbA},  \InPhase^{\SymbB} ,\SymbolTime{s})
		\quad \mathrm{s.t.} \quad (\ref{eq:Corr_constraint})
		.
	\end{aligned}
\end{equation}

%%%%%%%%%%%%%%%%%%%%%%%%%%%%%%%%%%%%%%%%%%%%%%%%%%%%%%%%%%%%%%%%%%%%%%%%%%%%%%%%%%%%%%%%%%%%%%%%%%%%%%%%%%%%%%%%%%%%%%%%%%%%%%%%%%%%%%%
%%%%%%%%%%%%%%%%%%%%%%%%%%%%%%%%%%%%%%%%%%%%%%%%%%%%%%%%%%%%%%%%%%%%%%%%%%%%%%%%%%%%%%%%%%%%%%%%%%%%%%%%%%%%%%%%%%%%%%%%%%%%%%%%%%%%%%%
%%%%%%%%%%%%%%%%%%%%%%%%%%%%%%%%%%%%%%%%%%%%%%%%%%%%%%%%%%%%%%%%%%%%%%%%%%%%%%%%%%%%%%%%%%%%%%%%%%%%%%%%%%%%%%%%%%%%%%%%%%%%%%%%%%%%%%%
\section{Proposed design method \label{sec_method}}

This section presents solutions to the main problem, which is how to simultaneously design $\RefWave(t)$, $\CarrierSig(t)$, $\InPhase^{\SymbA}$, and $\InPhase^{\SymbB}$.
A difficulty is that the performance index $\ObjectiveFunc(\RefWave, \CarrierSig, \InPhase^{\SymbA},  \InPhase^{\SymbB} ,\SymbolTime{s})$ is not given as an analytical form in (\ref{eq:def_objective}).
To address this difficulty, we focus on the decompositions of vibrations which is described in Section \ref{sec_factorize}.
The decomposition method transforms the performance index into a tractable form for the design.
Using the transformed performance index, Section \ref{sec_simple_design} presents design methods for $\RefWave(t)$, $\CarrierSig(t)$, $\InPhase^{\SymbA}$, and $\InPhase^{\SymbB}$ to simplify the structure of the receiver.

%%%%%%%%%%%%%%%%%%%%%%%%%%%%%%%%%%%%%%%%%%%%%%%%%%%%%%%%%%%%%%%%%%%%%%%%%%%%%%%%%%%%%%%%%%%%%%%%%%%%%%%%%%%%%%%%%%%%%%%%%%%%%%%%%%%%%%%
%%%%%%%%%%%%%%%%%%%%%%%%%%%%%%%%%%%%%%%%%%%%%%%%%%%%%%%%%%%%%%%%%%%%%%%%%%%%%%%%%%%%%%%%%%%%%%%%%%%%%%%%%%%%%%%%%%%%%%%%%%%%%%%%%%%%%%%
%%%%%%%%%%%%%%%%%%%%%%%%%%%%%%%%%%%%%%%%%%%%%%%%%%%%%%%%%%%%%%%%%%%%%%%%%%%%%%%%%%%%%%%%%%%%%%%%%%%%%%%%%%%%%%%%%%%%%%%%%%%%%%%%%%%%%%%
\subsection{Factorization of the objective fucntion \label{sec_factorize}}

\newcommand{\DefWave}{E}

\newcommand{\SSNDisp}[2]{x_{#1}(#2)}

\newcommand{\MagCoef}{\tilde{\Mag}}

This subsection factorizes the performance index $\ObjectiveFunc(\RefWave, \CarrierSig, \InPhase^{\SymbA},  \InPhase^{\SymbB} ,\SymbolTime{s})$ defined in (\ref{eq:def_objective}).
The following definition is used.

%%%%%%%%%%%%%%%%%%%%%%%%%%%%%%%%%%%%%%%%%%%%%%%%%%%%%%%%%%%%%%%%%%%%%%%%%%%%%%%%%%%%%%%%%%%%%%%%%%%%%%%%%%
\begin{definition}\label{def_SStoWave}
	For a given wave $\DefWave(t)$, $\Disp_{\DefWave}(t)$ is the steady state solution (particular solution) to the motion equation with respect to $\tilde{\Disp}(t)$:
	\begin{align}
	\CNTmass \frac{\mathrm{d}^{2}{\tilde{\Disp}(t)}}{\mathrm{d} t^{2}} 
	+ \viscosity \frac{\mathrm{d}{\tilde{\Disp}(t)}}{\mathrm{d} t} 
	+ \elasticity \tilde{\Disp}(t) & =  \charge \DefWave(t)
	.
	\end{align}
\end{definition}
%%%%%%%%%%%%%%%%%%%%%%%%%%%%%%%%%%%%%%%%%%%%%%%%%%%%%%%%%%%%%%%%%%%%%%%%%%%%%%%%%%%%%%%%%%%%%%%%%%%%%%%%%%

Under Definition \ref{def_SStoWave}, the solution to the motion equation (\ref{eq:CNT_motion}) is decomposed into $\Disp(t)=\Disp_{\mathrm{tr}}(t)+\Disp_{\InWave}(t)-\Disp_{\RefWave}(t)$ if $\Disp_{\InWave}(t)$ and $\Disp_{\RefWave}(t)$ exist. 
Here, $\Disp_{\mathrm{tr}}(t)$ is a transient component, which is decayed due to the viscosity.
By taking a sufficiently large symbol duration $\SymbolTime{s}$, $\Disp_{\mathrm{tr}}(t)$ can be negligible in the performance index $\ObjectiveFunc(\RefWave, \CarrierSig, \InPhase^{\SymbA},  \InPhase^{\SymbB} ,\SymbolTime{s})$ in (\ref{eq:def_objective}).
Therefore, we employ the approximation:
\begin{equation} 
\begin{aligned}	
\ObjectiveFunc(\RefWave, \CarrierSig, \InPhase^{\SymbA},  \InPhase^{\SymbB} ,\SymbolTime{s})
\approx
\SSObjectiveFunc(\RefWave, \CarrierSig, \InPhase^{\SymbA},  \InPhase^{\SymbB})
,
\end{aligned}
\end{equation}	
where $\SSObjectiveFunc(\RefWave, \CarrierSig, \InPhase^{\SymbA},  \InPhase^{\SymbB})$ consists of the steady state components, which is factorized as follows
\begin{equation} \label{eq:objective_factorization}
\begin{aligned}
%&
\SSObjectiveFunc(\RefWave, \CarrierSig, \InPhase^{\SymbA},  \InPhase^{\SymbB})
&:=
\Big|
\frac{\LinearCurrent}{\SymbolTime{1}}
\int_{0}^{\SymbolTime{1}}
(\SSDisp^{\SymbA}(t) + \SSDisp^{\SymbB}(t)) 
(\SSDisp^{\SymbA}(t) - \SSDisp^{\SymbB}(t)) \CarrierSig(t)
\mathrm{d} t
\Big|
\\&\;
=
\Big|
\frac{\LinearCurrent}{\SymbolTime{1}}
\int_{0}^{\SymbolTime{1}}
(
-
2 \SSNDisp{\RefWave}{t}
+
\SSNDisp{\InWave^{\SymbA}+\InWave^{\SymbB}}{t}
)
\SSNDisp{\InWave^{\SymbA}-\InWave^{\SymbB}}{t} \CarrierSig(t)
\mathrm{d} t
\Big|
\\&\;
=
\Big|
\frac{\LinearCurrent}{\SymbolTime{1}}
\int_{0}^{\SymbolTime{1}}
(
2\SSNDisp{\InWave^{\SymbB}}{t}
-
2 \SSNDisp{\RefWave}{t}
+
\SSNDisp{\InWave^{\SymbA}-\InWave^{\SymbB}}{t}
)
\SSNDisp{\InWave^{\SymbA}-\InWave^{\SymbB}}{t} \CarrierSig(t)
\mathrm{d} t
\Big|
,
\end{aligned}
\end{equation}	
where $\SSDisp^{\SymbA}(t)$ and $\SSDisp^{\SymbB}(t)$ are the steady state solutions to the motion equation (\ref{eq:CNT_motion}) corresponding to $\InPhase^{\SymbA}(t)$ and $\InPhase^{\SymbB}(t)$, respectively.	
The steady state solutions $\SSDisp^{\SymbA}(t)$ and $\SSDisp^{\SymbB}(t)$ are explicitly solved if the reference wave $\RefWave(t)$ is a specific periodic function such as a cosine wave.
The performance index $\SSObjectiveFunc(\RefWave, \CarrierSig, \InPhase^{\SymbA},  \InPhase^{\SymbB})$ can be represented as an explicit function if the carrier signal $\CarrierSig(t)$ is a specific periodic function such as a cosine wave.
Their details are described in Section \ref{sec_simple_design}.

%%%%%%%%%%%%%%%%%%%%%%%%%%%%%%%%%%%%%%%%%%%%%%%%%%%%%%%%%%%%%%%%%%%%%%%%%%%%%%%%%%%%%%%%%%%%%%%%%%%%%%%%%%%%%%%%%%%%%%%%%%%%%%%%%%%%%%%
%%%%%%%%%%%%%%%%%%%%%%%%%%%%%%%%%%%%%%%%%%%%%%%%%%%%%%%%%%%%%%%%%%%%%%%%%%%%%%%%%%%%%%%%%%%%%%%%%%%%%%%%%%%%%%%%%%%%%%%%%%%%%%%%%%%%%%%
%%%%%%%%%%%%%%%%%%%%%%%%%%%%%%%%%%%%%%%%%%%%%%%%%%%%%%%%%%%%%%%%%%%%%%%%%%%%%%%%%%%%%%%%%%%%%%%%%%%%%%%%%%%%%%%%%%%%%%%%%%%%%%%%%%%%%%%
\subsection{Design of the reference wave and the carrier signal for simple structures \label{sec_simple_design}}

On the basis of the factorized performance index $\SSObjectiveFunc(\RefWave, \CarrierSig, \InPhase^{\SymbA},  \InPhase^{\SymbB})$ in (\ref{eq:objective_factorization}), this subsection finds pairs of the reference wave $\RefWave(t)$ and the carrier signal $\CarrierSig(t)$ such that the structure of the receiver is simplified.
Let us define a coefficient for brief notation:
\begin{equation} \label{eq:def_MagCoef}
	\begin{aligned}
		%&
		\MagCoef
		:=
		\frac{ \charge \InMag }
		{\sqrt{(\elasticity - \CNTmass \InFreq^{2} )^{2} + (\viscosity \InFreq)^{2}}}
		.
	\end{aligned}
\end{equation}	
The main results are described in Theorems \ref{thm:wt_Corr} and \ref{thm:wt_RefWave}.

\newcommand{\coefRefWaveDesign}{\eta}

%%%%%%%%%%%%%%%%%%%%%%%%%%%%%%%%%%%%%%%%%%%%%%%%%%%%%%%%%%%%%%%%%%%%%%%%%%%%%%%%%%%%%%%%%%%%%%%%%%%%%%%%%%
\iffalse \subsubsection{A system without any carrier signal} \fi
\begin{theorem}[A system without any carrier signal] \label{thm:wt_Corr}
	Suppose that $\CarrierSig(t)=1$ holds. 
	If the condition
	\begin{equation} \label{eq:wt_correlator_cond}
		\begin{aligned}
			%&
			\RefWave(t) 
			=  - \coefRefWaveDesign  \InWave^{\SymbA}(t) + (1+\coefRefWaveDesign) \InWave^{\SymbB}(t)
			,
		\end{aligned}
	\end{equation}	
	holds, then the performance index $\SSObjectiveFunc(\RefWave, \CarrierSig, \InPhase^{\SymbA},  \InPhase^{\SymbB})$ is given by
	\begin{equation}  \label{eq:objective_result_wt_Corr}
		\begin{aligned}
			%&
			\SSObjectiveFunc(\RefWave, \CarrierSig, \InPhase^{\SymbA},  \InPhase^{\SymbB})
			=
			|
			\LinearCurrent
			\MagCoef^{2}
			(2 \coefRefWaveDesign +1)|( 1- \cos (\InPhase^{\SymbB}-\InPhase^{\SymbA} ) )
			.
		\end{aligned}
	\end{equation}			
\end{theorem}
%%%%%%%%%%%%%%%%%%%%%%%%%%%%%%%%%%%%%%%%%%%%%%%%%%%%%%%%%%%%%%%%%%%%%%%%%%%%%%%%%%%%%%%%%%%%%%%%%%%%%%%%%%
\textit{Proof.}	The proof is given in Appendix \ref{pf:wt_Corr}.	
%%%%%%%%%%%%%%%%%%%%%%%%%%%%%%%%%%%%%%%%%%%%%%%%%%%%%%%%%%%%%%%%%%%%%%%%%%%%%%%%%%%%%%%%%%%%%%%%%%%%%%%%%%
\begin{remark}
	The assumption $\CarrierSig(t)=1$ indicates that there is no carrier signal.
	The performance index $\SSObjectiveFunc(\RefWave, \CarrierSig, \InPhase^{\SymbA},  \InPhase^{\SymbB})$ is maximized with respect to $\InPhase^{\SymbA}$ and $\InPhase^{\SymbB}$ if the condition 
	\begin{align}
		%&
		\InPhase^{\SymbB}-\InPhase^{\SymbA} = \pi 
		,\label{eq:best_InPhase_wt_Corr}
	\end{align}
	holds.
	Note that for all $t$, $\SSDisp^{\SymbA}(t)=0$ and $\SSDisp^{\SymbB}(t)=0$ hold for $\coefRefWaveDesign=-1$ and  $\coefRefWaveDesign=0$, respectively. 
\end{remark}
%%%%%%%%%%%%%%%%%%%%%%%%%%%%%%%%%%%%%%%%%%%%%%%%%%%%%%%%%%%%%%%%%%%%%%%%%%%%%%%%%%%%%%%%%%%%%%%%%%%%%%%%%%

\newcommand{\CorrPhase}{\Phase_{\mathrm{c}}}

%%%%%%%%%%%%%%%%%%%%%%%%%%%%%%%%%%%%%%%%%%%%%%%%%%%%%%%%%%%%%%%%%%%%%%%%%%%%%%%%%%%%%%%%%%%%%%%%%%%%%%%%%%
\iffalse \subsubsection{A system without any reference wave} \fi
\begin{theorem}[A system without any reference wave] \label{thm:wt_RefWave}
	Supposing that $\RefWave(t)=0$ holds.
	For a given $\CorrPhase$, if the conditions
	\begin{align}
		\CarrierSig(t) &= \sqrt{2} \sin( 2 \InFreq t + \CorrPhase  )
		,\label{eq:wt_RefWave_condA}
		\\
		\InPhase^{\SymbA} &= - \InPhase^{\SymbB}
		,\label{eq:wt_RefWave_condB}
	\end{align}	
	holds, then the performance index $\SSObjectiveFunc(\RefWave, \CarrierSig, \InPhase^{\SymbA},  \InPhase^{\SymbB})$ is given by
	\begin{equation}  \label{eq:objective_result_wt_RefWave}
		\begin{aligned}
			%&
			\SSObjectiveFunc(\RefWave, \CarrierSig, \InPhase^{\SymbA},  \InPhase^{\SymbB})
			=
			\frac{
				|
				\LinearCurrent \MagCoef^{2}
				\sin ( 2 \InPhase^{\SymbB} ) 
				\cos (\CorrPhase - 2 \SSDispPhase ) 
				|
			}{\sqrt{2}}
			,
		\end{aligned}
	\end{equation}
	where
	\begin{align}
		\SSDispPhase
		:=
		-
		\arctan
		\frac{ \viscosity\InFreq  }{
			\elasticity - \CNTmass \InFreq^{2}
		}
		. \label{eq:def_phase_ss_solutions}
	\end{align}	
\end{theorem}
%%%%%%%%%%%%%%%%%%%%%%%%%%%%%%%%%%%%%%%%%%%%%%%%%%%%%%%%%%%%%%%%%%%%%%%%%%%%%%%%%%%%%%%%%%%%%%%%%%%%%%%%%%
\textit{Proof.}	The proof is given in Appendix \ref{pf:wt_RefWave}.	
%%%%%%%%%%%%%%%%%%%%%%%%%%%%%%%%%%%%%%%%%%%%%%%%%%%%%%%%%%%%%%%%%%%%%%%%%%%%%%%%%%%%%%%%%%%%%%%%%%%%%%%%%%
\begin{remark}
	The assumption $\RefWave(t)=0$ indicates that there is no reference wave.
	The performance index $\SSObjectiveFunc(\RefWave, \CarrierSig, \InPhase^{\SymbA},  \InPhase^{\SymbB})$ is maximized with respect to $\InPhase^{\SymbA}$, $\InPhase^{\SymbB}$, and $\CorrPhase$ if the conditions 
	\begin{align}
		%&
		\InPhase^{\SymbB} & = - \InPhase^{\SymbA} =  \pi/4 
		,\label{eq:best_InPhase_wt_RefWave}
		\\
		\CorrPhase & = 2 \SSDispPhase
		,\label{eq:best_CorrPhase_wt_RefWave}
	\end{align}
	hold.
\end{remark}
%%%%%%%%%%%%%%%%%%%%%%%%%%%%%%%%%%%%%%%%%%%%%%%%%%%%%%%%%%%%%%%%%%%%%%%%%%%%%%%%%%%%%%%%%%%%%%%%%%%%%%%%%%

Designing the receiver based on Theorems \ref{thm:wt_Corr} and \ref{thm:wt_RefWave} simplifies the system structure, which does not need the reference wave or the carrier signal as shown in Fig. \ref{fig:sys_config}.

%%%%%%%%%%%%%%%%%%%%%%%%%%%%%%%%%%%%%%%%%%%%%%%%%%%%%%%%%%%%%%%%%%%%%%%%%%%%%%%%%%%%%%%%%%%%%%%%%%%%%%%%%%%%%%%%%%%%%%%%%%%%%%%%%%%%%%%
%%%%%%%%%%%%%%%%%%%%%%%%%%%%%%%%%%%%%%%%%%%%%%%%%%%%%%%%%%%%%%%%%%%%%%%%%%%%%%%%%%%%%%%%%%%%%%%%%%%%%%%%%%%%%%%%%%%%%%%%%%%%%%%%%%%%%%%
%%%%%%%%%%%%%%%%%%%%%%%%%%%%%%%%%%%%%%%%%%%%%%%%%%%%%%%%%%%%%%%%%%%%%%%%%%%%%%%%%%%%%%%%%%%%%%%%%%%%%%%%%%%%%%%%%%%%%%%%%%%%%%%%%%%%%%%
\section{Conclusion \label{sec_conclusion}}

We proposed a design method to simplify the nanoscale receiver in structure with enhancing the performance for the phase detection.
The distance between constellation points is factorized, which is regarded as the performance measure.
Analyzing the factorized form yields the conditions that either the carrier signal or the reference wave is not employed.
The distance between constellation points is approximately maximized under the conditions.
No carrier signal or no reference wave is then employed, enhancing the performance for the phase detection.

%%%%%%%%%%%%%%%%%%%%%%%%%%%%%%%%%%%%%%%%%%%%%%%%%%%%%%%%%%%%%%%%%%%%%%%%%%%%%%%%%%%%%%%%%%%%%%%%%%%%%%%%%%%%%%%%%%%%%%%%%%%%%%%%%%%%%%%
%%%%%%%%%%%%%%%%%%%%%%%%%%%%%%%%%%%%%%%%%%%%%%%%%%%%%%%%%%%%%%%%%%%%%%%%%%%%%%%%%%%%%%%%%%%%%%%%%%%%%%%%%%%%%%%%%%%%%%%%%%%%%%%%%%%%%%%
%%%%%%%%%%%%%%%%%%%%%%%%%%%%%%%%%%%%%%%%%%%%%%%%%%%%%%%%%%%%%%%%%%%%%%%%%%%%%%%%%%%%%%%%%%%%%%%%%%%%%%%%%%%%%%%%%%%%%%%%%%%%%%%%%%%%%%%

\appendix

%%%%%%%%%%%%%%%%%%%%%%%%%%%%%%%%%%%%%%%%%%%%%%%%%%%%%%%%%%%%%%%%%%%%%%%%%%%%%%%%%%%%%%%%%%%%%%%%%%%%%%%%%%%%%%%%%%%%%%%%%%%%%%%%%%%%%%%
%%%%%%%%%%%%%%%%%%%%%%%%%%%%%%%%%%%%%%%%%%%%%%%%%%%%%%%%%%%%%%%%%%%%%%%%%%%%%%%%%%%%%%%%%%%%%%%%%%%%%%%%%%%%%%%%%%%%%%%%%%%%%%%%%%%%%%%
%%%%%%%%%%%%%%%%%%%%%%%%%%%%%%%%%%%%%%%%%%%%%%%%%%%%%%%%%%%%%%%%%%%%%%%%%%%%%%%%%%%%%%%%%%%%%%%%%%%%%%%%%%%%%%%%%%%%%%%%%%%%%%%%%%%%%%%
\section{Proof of Theorem \ref{thm:wt_Corr} \label{pf:wt_Corr}}

\newcommand{\PhaseTempA}{\theta_{1}}
\newcommand{\PhaseTempB}{\theta_{2}}
Substituting $\CarrierSig(t)=1$ and (\ref{eq:wt_correlator_cond}) into the performance index $\SSObjectiveFunc(\RefWave, \CarrierSig, \InPhase^{\SymbA},  \InPhase^{\SymbB})$ in (\ref{eq:objective_factorization}) yields
\begin{equation} \label{eq:objective_factorization_wt_Corr}
\begin{aligned}
%&
\SSObjectiveFunc(\RefWave, \CarrierSig, \InPhase^{\SymbA},  \InPhase^{\SymbB})
&=
\Big|
\frac{\LinearCurrent}{\SymbolTime{1}}
\int_{0}^{\SymbolTime{1}}
(2\coefRefWaveDesign+1)
\SSNDisp{\InWave^{\SymbA}-\InWave^{\SymbB}}{t}^{2}
\mathrm{d} t
\Big|
=
\frac{|\LinearCurrent(2\coefRefWaveDesign+1)|}{\SymbolTime{1}}
\int_{0}^{\SymbolTime{1}}
\SSNDisp{\InWave^{\SymbA}-\InWave^{\SymbB}}{t}^{2}
\mathrm{d} t
.
\end{aligned}
\end{equation}
The relation 
\begin{equation} 
\begin{aligned}
\InWave^{\SymbA}(t)-\InWave^{\SymbB}(t) 
&= \InMag \cos ( \InFreq t + \InPhase^{\SymbA} ) - \InMag \cos ( \InFreq t + \InPhase^{\SymbB} )
\\&=  \InMag \sqrt{ 2 - 2  \cos (\InPhase^{\SymbB}-\InPhase^{\SymbA} ) } \cos(\InFreq t + \PhaseTempA )
,
\end{aligned}
\end{equation}
holds for some $\PhaseTempA$.
Using this relation, for some $\PhaseTempB$, the steady state solution $\SSNDisp{\InWave^{\SymbA}-\InWave^{\SymbB}}{t}$ is described as follows:
\begin{align}
%&
\SSNDisp{\InWave^{\SymbA}-\InWave^{\SymbB}}{t}
&=
\sqrt{ 2 - 2  \cos (\InPhase^{\SymbB}-\InPhase^{\SymbA} ) }
\MagCoef
\cos ( \InFreq t + \PhaseTempB  ) 
.\label{eq:SSDisp_wt_Corr}
\end{align}
Thus, substituting (\ref{eq:SSDisp_wt_Corr}) into (\ref{eq:objective_factorization_wt_Corr}) yields (\ref{eq:objective_result_wt_Corr}) because $\int_{0}^{\SymbolTime{1}} \cos(\InFreq t + \PhaseTempB)^{2} \mathrm{d} t = \SymbolTime{1}/2$ holds for any $\PhaseTempB$.
This completes the proof.	
\hfill$\square$

%%%%%%%%%%%%%%%%%%%%%%%%%%%%%%%%%%%%%%%%%%%%%%%%%%%%%%%%%%%%%%%%%%%%%%%%%%%%%%%%%%%%%%%%%%%%%%%%%%%%%%%%%%%%%%%%%%%%%%%%%%%%%%%%%%%%%%%
%%%%%%%%%%%%%%%%%%%%%%%%%%%%%%%%%%%%%%%%%%%%%%%%%%%%%%%%%%%%%%%%%%%%%%%%%%%%%%%%%%%%%%%%%%%%%%%%%%%%%%%%%%%%%%%%%%%%%%%%%%%%%%%%%%%%%%%
%%%%%%%%%%%%%%%%%%%%%%%%%%%%%%%%%%%%%%%%%%%%%%%%%%%%%%%%%%%%%%%%%%%%%%%%%%%%%%%%%%%%%%%%%%%%%%%%%%%%%%%%%%%%%%%%%%%%%%%%%%%%%%%%%%%%%%%
\section{Proof of Theorem \ref{thm:wt_RefWave} \label{pf:wt_RefWave}}

The relations $\InWave^{\SymbA}(t) +  \InWave^{\SymbB}(t) = \InMag \cos ( \InFreq t  - \InPhase^{\SymbB} )  + \InMag \cos ( \InFreq t  + \InPhase^{\SymbB})  = 2 \InMag \cos ( \InFreq t  )  \cos ( \InPhase^{\SymbB} )$ and $\InWave^{\SymbA}(t) -  \InWave^{\SymbB}(t) = \InMag \cos ( \InFreq t  - \InPhase^{\SymbB} ) - \InMag \cos ( \InFreq t  + \InPhase^{\SymbB}) = 2 \InMag \sin ( \InFreq t  )  \sin ( \InPhase^{\SymbB} )$ hold because of the condition (\ref{eq:wt_RefWave_condB}).
The steady state solutions with respect to $\InWave^{\SymbA}(t) \pm \InWave^{\SymbB}(t)$ are given as
\begin{align}
& 
\SSNDisp{\InWave^{\SymbA}+\InWave^{\SymbB}}{t}
=
2 \MagCoef  \cos ( \InPhase^{\SymbB} )  
\cos ( \InFreq t + \SSestDispPhase  ) 
,\label{eq:SSDisp_wt_RefWave_p} \\&
\SSNDisp{\InWave^{\SymbA}-\InWave^{\SymbB}}{t}
=
2 \MagCoef \sin ( \InPhase^{\SymbB} )  
\sin ( \InFreq t + \SSestDispPhase    ) 
,\label{eq:SSDisp_wt_RefWave_m}
\end{align}	
where $\SSestDispPhase \in \{ \SSDispPhase, \SSDispPhase-\pi \}$.
Substituting $\RefWave(t)=0$, (\ref{eq:SSDisp_wt_RefWave_p}), and (\ref{eq:SSDisp_wt_RefWave_m}) into (\ref{eq:objective_factorization}) yields 
\begin{equation} \label{eq:objective_factorization_wt_RefWave}
\begin{aligned}
%&
\SSObjectiveFunc(\RefWave, \CarrierSig, \InPhase^{\SymbA},  \InPhase^{\SymbB})
&=
\Big|
\frac{\LinearCurrent}{\SymbolTime{1}}
\int_{0}^{\SymbolTime{1}}
\SSNDisp{\InWave^{\SymbA}+\InWave^{\SymbB}}{t}
\SSNDisp{\InWave^{\SymbA}-\InWave^{\SymbB}}{t} \CarrierSig(t)
\mathrm{d} t
\Big|
\\&
=
\Big|
\frac{
	\LinearCurrent \MagCoef^{2}
}
{ \SymbolTime{1}}
\int_{0}^{\SymbolTime{1}}
4 \cos ( \InPhase^{\SymbB} ) \sin ( \InPhase^{\SymbB} )
\cos ( \InFreq t + \SSestDispPhase  ) 
\sin ( \InFreq t + \SSestDispPhase  ) 
\CarrierSig(t)
\mathrm{d} t
\Big|
\\&
=
\Big|
\frac{
	\LinearCurrent \MagCoef^{2}
	\sin ( 2 \InPhase^{\SymbB} )
}
{ \SymbolTime{1}}
\int_{0}^{\SymbolTime{1}}
\sin ( 2 \InFreq t + 2 \SSestDispPhase  ) 
\CarrierSig(t)
\mathrm{d} t
\Big|
\\&
=
\Big|
\frac{
	\LinearCurrent \MagCoef^{2}
	\sin ( 2 \InPhase^{\SymbB} )
}
{ \SymbolTime{1}}
\int_{0}^{\SymbolTime{1}}
\sin ( 2 \InFreq t + 2 \SSDispPhase  ) 
\CarrierSig(t)
\mathrm{d} t
\Big|
.
\end{aligned}
\end{equation}
Meanwhile, the carrier signal $\CarrierSig(t)$ is written by 
\begin{equation} \label{eq:correlator_trans}
\begin{aligned}	
\CarrierSig(t)
&=\sqrt{2}  \sin ( 2 \InFreq t + 2 \SSDispPhase + (\CorrPhase - 2 \SSDispPhase)  ) 
\\&
=
\sqrt{2}  \sin ( 2 \InFreq t + 2 \SSDispPhase ) \cos (\CorrPhase - 2 \SSDispPhase ) 
+
\sqrt{2}  \cos ( 2 \InFreq t + 2 \SSDispPhase ) \sin (\CorrPhase - 2 \SSDispPhase ) 
.
\end{aligned}
\end{equation}
Substituting (\ref{eq:correlator_trans}) into (\ref{eq:objective_factorization_wt_RefWave}) yields  (\ref{eq:objective_result_wt_RefWave}) because $\int_{0}^{\SymbolTime{1}} \sin( 2 \InFreq t + 2 \SSDispPhase )^{2} \mathrm{d} t = \SymbolTime{1}/2$ and  $\int_{0}^{\SymbolTime{1}} $ $\sin( 2 \InFreq t + 2 \SSDispPhase )\cos( 2 \InFreq t + 2 \SSDispPhase ) \mathrm{d} t = 0$ holds.
This completes the proof.
\hfill$\square$

%%%%%%%%%%%%%%%%%%%%%%%%%%%%%%%%%%%%%%%%%%%%%%%%%%%%%%%%%%%%%%%%%%%%%%%%%%%%%%%%%%%%%%%%%%%%%%%%%%%%%%%%%%%%%%%%%%%%%%%%%%%%%%%%%%%%%%%
%%%%%%%%%%%%%%%%%%%%%%%%%%%%%%%%%%%%%%%%%%%%%%%%%%%%%%%%%%%%%%%%%%%%%%%%%%%%%%%%%%%%%%%%%%%%%%%%%%%%%%%%%%%%%%%%%%%%%%%%%%%%%%%%%%%%%%%
%%%%%%%%%%%%%%%%%%%%%%%%%%%%%%%%%%%%%%%%%%%%%%%%%%%%%%%%%%%%%%%%%%%%%%%%%%%%%%%%%%%%%%%%%%%%%%%%%%%%%%%%%%%%%%%%%%%%%%%%%%%%%%%%%%%%%%%

%\bibliography{CNT_simplified_structure_Yuji_Ito_20190125bib.bib}

%%%%%%%%%%%%%%%%%%%%%%%%%%%%%%%%%%%%%%%%%%%%%%%%%%%%%%%%%%%%%%%%%%%%%%%%%%%%%%%%%%%%%%%%%%%%%%%%%%%%%%%%%%%%%%%%%%%%%%%%%%%%%%%%%%%%%%%
%%%%%%%%%%%%%%%%%%%%%%%%%%%%%%%%%%%%%%%%%%%%%%%%%%%%%%%%%%%%%%%%%%%%%%%%%%%%%%%%%%%%%%%%%%%%%%%%%%%%%%%%%%%%%%%%%%%%%%%%%%%%%%%%%%%%%%%
%%%%%%%%%%%%%%%%%%%%%%%%%%%%%%%%%%%%%%%%%%%%%%%%%%%%%%%%%%%%%%%%%%%%%%%%%%%%%%%%%%%%%%%%%%%%%%%%%%%%%%%%%%%%%%%%%%%%%%%%%%%%%%%%%%%%%%%

\end{document}